\documentclass[nofootinbib,prd,twocolumn,showpacs,showkeys,preprintnumbers]{revtex4-1}
\usepackage{hyperref,amssymb,amsmath,mathrsfs,bm,graphicx}
\begin{document}
\title {Complexity of the Bondi metric}
\author{L. Herrera}
\email{lherrera@usal.es}
\affiliation{Instituto Universitario de F\'isica
Fundamental y Matem\'aticas, Universidad de Salamanca, Salamanca 37007, Spain }
\author{A. Di Prisco}
\email{adiprisc@ciens.ucv.ve}
\affiliation{Escuela de F\'\i sica, Facultad de Ciencias, Universidad Central de Venezuela, Caracas 1050, Venezuela}
\author{J.  Carot}
\email{jcarot@uib.cat}
\affiliation{Departament de  F\'{\i}sica, Universitat Illes Balears, E-07122 Palma de Mallorca, Spain}

\begin{abstract}
A recently introduced concept of complexity for relativistic fluids is extended to the vacuum solutions represented by the Bondi metric. A complexity hierarchy is established, ranging from the Minkowski spacetime (the simplest one) to gravitationally radiating systems (the more complex). Particularly interesting is  the possibility to differentiate between natural non--radiative (NNRS) and non--natural non--radiative (NNNRS) systems, the latter appearing to be simpler than the former. The relationship between vorticity and the degree of complexity is stressed.
\end{abstract}
\date{\today}
\maketitle
\section{Introduction}
In a recent series of papers we have introduced a new concept of complexity for self--gravitating relativistic fluids, and have applied it to the spherically symmetric case (both in the static \cite{css} and the dynamic situation \cite{csd}) and to the axially symmetric static case \cite{cax}. Applications of this concept   to other theories of gravity have been proposed in  \cite{ot1, ot2}, while  the charged case  has been considered in \cite{csd} and \cite{ch}. Also, applications   for some particular cases of cylindrically symmetric fluid distributions, may be found in \cite{cil}.

Our purpose in this paper consists in extending the above mentioned concept of complexity to vacuum spacetimes. More specifically we shall consider the Bondi metric \cite{Boal}, which includes the Minkowski spacetime, the static Weyl metrics, non--radiative non--static metrics and gravitationally radiating metrics.
Besides the fact that the Bondi metric covers a vast numbers of spacetimes, it has, among other
things, the virtue of providing a clear and precise criterion for the existence of gravitational
radiation. Namely, if the news function is zero over a time interval, then there
is no radiation over that interval. 

In the case of fluid distributions the variable(s) measuring the complexity of the fluid (the complexity factor(s))  appear in the trace free part of the orthogonal splitting of the electric  Riemann tensor \cite{11p, 12p, 14p, 9}. In vacuum the Riemann tensor and the Weyl tensor are the same, so we shall start by calculating the scalar functions defining the electric part of the Weyl tensor for the Bondi metric. Following the results obtained for the fluid case,  we shall consider the scalars defining this tensor as the complexity factors. Next we shall establish a hierarchy of  spacetimes according to their complexity. Particularly  appealing is the possibility to discriminate between  two classes of spacetimes that depend on time but are not radiative (vanishing of the news function).  These two classes were called by Bondi \cite{Boal} natural and non--natural non--radiative moving systems, and are characterized by different forms  of the mass aspect. As we shall see they exhibit different degree of complexity. 

Unfortunately, though, up to the leading order of the complexity factors analyzed here, it is  impossible to discriminate between different radiative systems according to their complexity. Higher order terms would be necessary for that purpose, although it is not clear at this point if is possible to establish such a hierarchy of radiative systems after all. 

Finally we emphasize the conspicuous link between vorticity and complexity, and discuss about some open issues in the last section.

\section{The Bondi's formalism}
The general form of an axially and reflection symmetric asymptotically flat
metric given by Bondi \cite{Boal}  is  (for the general case see \cite{Sachs})
\begin{eqnarray}
ds^2 & = & \left(\frac{V}{r} e^{2\beta} - U^2 r^2 e^{2\gamma}\right) du^2
+ 2 e^{2\beta} du dr \nonumber \\
& + & 2 U r^2 e^{2\gamma} du d\theta
- r^2 \left(e^{2 \gamma} d\theta^2 + e^{-2\gamma} \sin^2{\theta} 
d\phi^2\right),
\label{Bm}
\end{eqnarray}
where $V, \beta, U$ and $\gamma$ are functions of
$u, r$ and $\theta$.

We number the coordinates $x^{0,1,2,3} = u, r, \theta, \phi$ respectively.
$u$ is a timelike coordinate ($g_{uu}>0$ ) converging to the retarded time as $r\rightarrow \infty$. The hypersurfaces  $u=constant$ define  null surfaces (their normal vectors  are  null vectors), which at null infinity ($r\rightarrow \infty$) coincides with the Minkowski null light cone
open to the future. $r$ is a null coordinate ($g_{rr}=0$) and $\theta$ and
$\phi$ are two angle coordinates (see \cite{Boal} for details).

Regularity conditions in the neighborhood of the polar axis
($\sin{\theta}=0$), imply that
as $\sin{\theta} \rightarrow 0$
\begin{equation}
V, \beta, U/\sin{\theta}, \gamma/\sin^2{\theta},
\label{regularity}
\end{equation}
each equals a function of $\cos{\theta}$ regular on the polar axis.

The four metric functions are assumed to be expanded in series of $1/r$,
then using the field equations Bondi gets

\begin{equation}
\gamma = c r^{-1} + \left(C - \frac{1}{6} c^3\right) r^{-3}
+ ...,
\label{ga}
\end{equation}
\newpage
\begin{equation}
U = - \left(c_\theta + 2 c \cot{\theta}\right) r^{-2} + \left[2
N+3cc_{\theta}+4c^2 \cot{\theta}\right]r^{-3}...,
\label{U}
\end{equation}
\begin{widetext}
\begin{eqnarray}
V  = r - 2 M
 -  \left( N_\theta + N \cot{\theta} -
c_{\theta}^{2} - 4 c c_{\theta} \cot{\theta} -
\frac{1}{2} c^2 (1 + 8 \cot^2{\theta})\right) r^{-1} + ...,
\label{V}
\end{eqnarray}
\end{widetext}
\begin{equation}
\beta = - \frac{1}{4} c^2 r^{-2} + ...,
\label{be}
\end{equation}
 where $c$, $C$, $N$ and $M$ are functions of $u$ and $\theta$ satisfying the constraint

\begin{equation}
4C_u = 2 c^2 c_u + 2 c M + N \cot{\theta} - N_\theta,
\label{C}
\end{equation}
and letters as
subscripts denote derivatives.
The three functions $c, M$ and $N$ are further
related by the supplementary conditions
\begin{equation}
M_u = - c_u^2 + \frac{1}{2}
\left(c_{\theta\theta} + 3 c_{\theta} \cot{\theta} - 2 c\right)_u,
\label{Mass}
\end{equation}
\begin{equation}
- 3 N_u = M_\theta + 3 c c_{u\theta} + 4 c c_u \cot{\theta} + c_u c_\theta.
\label{N}
\end{equation}

In the static case $M$ equals the mass of the system and is called by Bondi the ``mass aspect'', whereas $N$ and $C$
are closely related to the dipole and quadrupole moments respectively.

Next, Bondi defines the mass $m(u)$ of the system as
\begin{equation}
m(u) = \frac{1}{2} \int_0^\pi{M \sin{\theta} d\theta},
\label{m}
\end{equation}
which by virtue of (\ref{Mass}) and (\ref{regularity}) yields
\begin{equation}
m_u = - \frac{1}{2} \int_0^\pi{c_u^2 \sin{\theta} d\theta}.
\label{muI}
\end{equation}

Let us now recall the main conclusions emerging from  Bondi's approach.
\begin{enumerate}
\item If $\gamma, M$ and $N$ are known for some $u=a$(constant) and
$c_u$ (the news function) is known for all $u$ in the interval
$a \leq u \leq b$,
then the system is fully determined in that interval. In other words,
whatever happens at the source, leading to changes in the field,
it can only do so by affecting $c_u$ and vice versa. In the
light of this comment the relationship between news function
and the occurrence of radiation becomes clear.
\item As it follows from (\ref{muI}), the mass of a system is constant
if and only if there is no news.
\end{enumerate}

Now, for an observer at rest in the frame of (\ref{Bm}), the four-velocity
vector has components
\begin{equation}
V^{\alpha} = \left(\frac{1}{A}, 0, 0, 0\right),
\label{fvct}
\end{equation}
with
\begin{equation}
A \equiv \left(\frac{V}{r} e^{2\beta} - U^2 r^2 e^{2\gamma}\right)^{1/2}.
\label{A}
\end{equation}

Next, let us  introduce the unit, spacelike vectors ${\bf K}$, ${\bf L}$, ${\bf S}$, with components
\begin{equation}
K^{\alpha} = \left(\frac{1}{A}, -e^{-2\beta}A, 0, 0\right)\quad L^{\alpha} = \left(0, Ure^{\gamma}e^{-2\beta}, -\frac{e^{-\gamma}}{r}, 0\right)
\label{K}
\end{equation}
\begin{equation}
S^{\alpha} = \left(0, 0, 0, -\frac{e^{\gamma}}{r\sin \theta} \right),
\label{K}
\end{equation}
or
\begin{equation}
V_{\alpha} = \left(A, \frac{e^{2\beta}}{A}, \frac{Ur^2e^{2\gamma}}{A}, 0\right),\quad K_{\alpha} = \left(0, \frac{e^{2\beta}}{A}, \frac{Ur^2e^{2\gamma}}{A}, 0\right),
\label{fvctc}
\end{equation}
\begin{equation}
L_{\alpha} = \left(0, 0, e^{\gamma}r, 0\right), \quad S_{\alpha} = \left(0, 0, 0, e^{-\gamma} r\sin \theta \right),
\label{K}
\end{equation}
satisfying  the following relations:
\begin{equation}
V_{\alpha} V^{\alpha}=-K^{\alpha} K_{\alpha}=-L^{\alpha} L_{\alpha}=-S^{\alpha} S_{\alpha}=1,
\label{4n}
\end{equation}
\begin{equation}
V_{\alpha} K^{\alpha}=V^{\alpha} L_{\alpha}=V^{\alpha} S_{\alpha}=K^{\alpha} L_{\alpha}=K^{\alpha} S_{\alpha}=S^{\alpha} L_{\alpha}=0.
\label{5n}
\end{equation}
The unitary vectors $V^\alpha, L^\alpha, S^\alpha, K^\alpha$ form a canonical  orthonormal tetrad ($e^{(a)}_\alpha$), such that  $$e^{(0)}_\alpha=V_\alpha,\quad e^{(1)}_\alpha=K_\alpha,\quad
e^{(2)}_\alpha=L_\alpha,\quad e^{(3)}_\alpha=S_\alpha,$$ with $a=0,\,1,\,2,\,3$ (latin indices labeling different vectors of the tetrad). The  dual vector tetrad $e_{(a)}^\alpha$  is easily computed from the condition 

$$ \eta_{(a)(b)}= g_{\alpha\beta} e_{(a)}^\alpha e_{(b)}^\beta,$$ where $ \eta_{(a)(b)}$ denotes the Minkowski metric.

For the observer defined by  (\ref{fvct}) the vorticity vector may be
written as (see \cite{1} for details)
\begin{equation}
\omega^\alpha = \left(0, 0, 0, \omega^{\phi}\right).
\label{oma}
\end{equation}
The explicit expressions  for $ \omega^{\phi}$ and its absolute value $\Omega  \equiv  \left(- \omega_\alpha \omega^\alpha\right)^{1/2}$ are given in the Appendix C.
\section{Complexity factors and electric and magnetic  parts of Weyl tensor}
As we mentioned in the Introduction we shall extend the definition of complexity introduced in \cite{css, csd} to the vacuum case, this implies considering  the scalars defining the electric Weyl  tensor  as the complexity factors. Besides the electric part of the Weyl tensor, we shall also use its magnetic part in the discussion, accordingly we shall calculate its corresponding scalars as well.

The electric and magnetic parts of Weyl tensor, $E_{\alpha \beta}$ and
$H_{\alpha\beta}$, respectively, are formed from the Weyl tensor $C_{\alpha
\beta \gamma \delta}$ and its dual
$\tilde C_{\alpha \beta \gamma \delta}$ by contraction with the four
velocity vector given by (\ref{fvct}):
\begin{equation}
E_{\alpha \beta}=C_{\alpha \gamma \beta \delta}V^{\gamma}V^{\delta},
\label{electric}
\end{equation}
\begin{eqnarray}
H_{\alpha \beta}&=&\tilde C_{\alpha \gamma \beta \delta}V^{\gamma}V^{\delta}=
\frac{1}{2}\epsilon_{\alpha \gamma \epsilon \delta} C^{\epsilon
\delta}_{\quad \beta \rho} V^{\gamma}
V^{\rho},\nonumber \\
&& \epsilon_{\alpha \beta \gamma \delta} \equiv \sqrt{-g}
\;\;\eta_{\alpha \beta \gamma \delta},
\label{magnetic}
\end{eqnarray}
where
$\eta_{\alpha\beta\gamma\delta}$ is the permutation symbol.

Also note that

$$  \sqrt{-g} = r^2 \sin \theta e^{2\beta} \approx r^2 \sin \theta
\exp{(-\frac{c^2}{2r^2})} \approx r^2 \sin \theta + O(1).$$

The electric part of the Weyl tensor has only three independent non-vanishing components, whereas only two components define the magnetic part. Thus  we may  write
\begin{widetext}
\begin{eqnarray}
E_{\alpha \beta}=\mathcal{E}_1\left(K_\alpha L_\beta+L_\alpha K_\beta\right)
+\mathcal{E}_2\left(K_\alpha K_\beta+\frac{1}{3}h_{\alpha \beta}\right)+\mathcal{E}_3\left(L_\alpha L_\beta+\frac{1}{3}h_{\alpha \beta}\right), \label{13}
\end{eqnarray}
\end{widetext}

and
\begin{equation}
H_{\alpha\beta}=H_1(S_\alpha K_\beta+S_\beta
K_\alpha)+H_2(S_\alpha L_\beta+S_\beta L_\alpha)\label{H'}.
\end{equation}

with $h_{\mu \nu}=g_{\mu\nu}-V_\nu V_\mu$, and

\begin{equation}
\mathcal{E}_1=L^\alpha K^\beta E_{\alpha \beta},
\label{ew3}
\end{equation}
\begin{equation}
\mathcal{E}_2=(2K^\alpha K^\beta+ L^\alpha L^\beta)E_{\alpha \beta},
\label{ew4}
\end{equation}
\begin{equation}
\mathcal{E}_3=(2L^\alpha L^\beta+ K^\alpha K^\beta)E_{\alpha \beta},
\label{ew4}
\end{equation}
these three scalars will be considered  the complexity  factors of  our solutions.

For the magnetic part we have
\begin{equation}
H_2=S^\alpha L^\beta H_{\alpha \beta},
\label{ew5}
\end{equation}
 \begin{equation}
H_1=S^\alpha K^\beta H_{\alpha \beta}.
\label{ew5}
\end{equation}
 
Explicit expressions for these scalars are given in the Appendixes A and B.

In \cite{HSC} it was obtained that if we put $H^{\alpha}_{\beta}=0$ then
the field is non--radiative and 
 up to order $1/r^3$ in $\gamma$, the metric is static, and the
mass, the ``dipole'' ($N$) and the ``quadrupole'' ($C$) moments correspond
to a static situation. However, the
time dependence might enter through coefficients of higher order in
$\gamma$, giving rise to what Bondi calls ``non--natural--non--radiative
moving system'' (NNNRS). In this latter case,
the system keeps the first three  moments independent of time, but allows
for time dependence of higher moments.  This class of solutions is characterized by $M_{\theta}=0$. 

A second family of time dependent non--radiative solutions exists for which $M_{\theta}\neq 0$. These are called natural non--radiative
moving system'' (NNRS), and their magnetic Weyl tensor is non--vanishing.

We are now ready to discuss the hierarchy of different spacetimes belonging to the Bondi family, according to their complexity.

\section{Hierarchy of complexity}
The simplest spacetime corresponds to the vanishing of the three complexity factors, and this is just Minkowski.

Indeed, as it was shown in \cite{HSC},  if we assume
$E^{\alpha}_{\beta}=0$ and use regularity conditions, we find that  the spacetime must be Minkowski,
giving further support to the
conjecture that there are no purely magnetic vacuum space--times 
\cite{Bonnor}.

On the other end (maximal complexity) we have a gravitationally radiating system which requires all three complexity factors to be different from zero.

Indeed, let us  assume that $\mathcal{E}_1=0$, then it follows at once from (\ref{e1}) that $c_u=0$ (otherwise $c_u$ would be a non--regular function of $\theta$ on the symmetry axis). Thus $\mathcal{E}_1=0$ implies
 that the system is non--radiative.
 
 If instead we assume that $\mathcal{E}_2=0$, then from the first order in (\ref{e2}) we obtain that $c_{uu}=0$, this implies that either $c_u=0$ or $c_u\sim u$. Bondi refers to this latter case as ``mass loss without radiative Riemann tensor'' and dismisses it as being of little physical significance. As a matter of fact, in this latter case the system would be radiating ``forever'', which  according to (\ref{muI}) requires an unbounded source, incompatible with an asymptotically flat spacetime. Thus in this case too, we have $c_u=0$, and the system is non--radiative.
 
 Finally, if we assume $\mathcal{E}_3=0$ it follows at once from the first order in (\ref{e3}), that $c_{uu}=0$, leading to $c_{u}=0$, according to the argument above.

Thus, a radiative system requires all  three complexity factors to be nonvanishing, implying  a maximal complexity.

In the middle of the two extreme cases   we have, on the one hand  the spherically symmetric spacetime (Schwarzschild), characterized by a single complexity factor (the same applies for any static metric), $\mathcal{E}_1=\mathcal{E}_3=0$, and $\mathcal{E}_2=\frac{3M}{r^3}$. On the other hand, we have the non--static non--radiative case.

Let us now analyze in detail this latter case. There are two subclasses in this group of solutions, which using Bondi notation are:
\begin{enumerate}
\item Natural--non--radiative systems (NNRS) characterized by $M_\theta\neq0$.
\item Non--natural--non--radiative systems (NNNRS) characterized by $M_\theta=0$.
\end{enumerate}

Let us first consider the NNNRS subcase. Using (\ref{e1}) we obtain  ${\cal E}_1=0$, (up to order $1/r^3$),  while the first non--vanishing terms in ${\cal E}_2$ and ${\cal E}_3$ are respectively,
$3M$ and $0$,
where (\ref{C}), (\ref{Mass}), (\ref{N}), (\ref{e2}) and (\ref{e3}) have been used. 

Thus,  the NNNRS are characterized by only one non--vanishing complexity factor ( ${\cal E}_2$).  Furthermore, as it follows from (\ref{Om2}) the vorticity of the congruence of observers at rest with respect to the frame of (\ref{Bm}) vanishes, and the field is purely electric. However as mentioned before we cannot conclude that the field is static, since the $u$ dependence might appear through coefficients of higher order in $\gamma$.

Let us now consider the  ``natural--non--radiative  system'' (NNRS). In this subcase, using (\ref{e1}) we obtain  ${\cal E}_1=0$, (up to order $1/r^3$) as for the NNNRS subcase, while the first non--vanishing term in ${\cal E}_2$ and ${\cal E}_3$  (up to order $1/r^3$)  are  respectively $3M+\frac{M_{\theta \theta}}{4}-\frac{M_{ \theta}\cot \theta}{4}$ and  $\frac{M_{\theta \theta}}{2}-\frac{M_{ \theta}\cot \theta}{2}$. 

Also, up to the same order, it follows from (\ref{h1}) and (\ref{h2}) that $H_1=0$ for both subcases, while the corresponding term in $H_2$ is (for NNRS)
\begin{equation}
-\frac{1}{4}(M_{\theta \theta}-M_\theta \cot\theta),
\label{h2r3}
\end{equation}
which of course vanishes for the NNNRS subcase.

It should be observed that if we assume  ${\cal E}_3=0$ or $H_2=0$  then it follows at once from the above that 

\begin{equation}
M_{\theta \theta}-M_\theta \cot\theta=0 \Rightarrow M=a\cos\theta, \quad a=constant.
\label{h2r3f}
\end{equation}

But this implies because of   (\ref{m}) that the Bondi mass function of the system vanishes. Therefore, the only physically meaningful NNRS requires ${\cal E}_3\neq0$, $\Omega\neq0$  and $H_2\neq0$  implying that the complexity is characterized by two complexity factors (${\cal E}_2$, ${\cal E}_3$).

\section{Conclusions}
We have seen so far that the extension of the concept of complexity, adopted for fluids in \cite{css, csd, cax}, may be extended to the vacuum case without much trouble and provides sensible results. 

The three complexity factors corresponding to the three scalars defining the electric part of the Weyl tensor, allow us  to establish a hierarchy of solutions according to their complexity. 

The simplest system (Minkowski) is characterized by the vanishing of all the complexity factors. Next, the static case (including Schwarzchild) is described by a single complexity factor. 

The time dependent non--radiative solutions split in two subgroups depending on the form of the mass aspect $M$. If $M_\theta=0$ which corresponds to the NNNRS the complexity is similar to the static case.  Also, in this case, as in the static situation, the vorticity vanishes and the field is purely electric. This result could  suggest that in fact  NNNRS are just static, and no time dependence appears  in the coefficients of higher order in
$\gamma$. On the contrary for the NNRS there are two complexity factors, the vorticity is non--vanishing and  the field is not purely electric. 

All these results are summarized  in Tables I and II.
Thus, NNNRS and NNRS are clearly differentiated through their degree of complexity, as measured by the complexity factors considered here. 

The fact that radiative systems necessarily decay into NNRS, NNNRS or static systems, since  the Bondi mass function must be finite, suggests that higher degrees of complexity might be  associated with stronger stability. Of course a proof of this conjecture requires a much more detailed analysis.

It is also worth mentioning  the conspicuous link between vorticity and complexity factors. Indeed vorticity appears only in NNRS and radiative systems, which are the most complex systems, while it is absent in the simplest systems (Minkowski, static, NNNRS). In the radiative case there are contributions at order $\mathcal{O}(r^{-1})$  related to the news function, and at  order $\mathcal{O}(r^{-2})$, while for  the NNRS there are only contributions at order $\mathcal{O}(r^{-2})$, these  describe the effect of the tail of the wave, thereby  providing  ``observational'' evidence for the violation of the Huygens's principle, a problem largely discussed in the literature (see for example\cite{Boal, tail1, BoNP, tail3, tail2, tail6, tail4, tail5} and references therein). 

We would like to conclude with two questions which, we believe, deserve further attention:
\begin{itemize}
\item Is it possible to further refine the scheme proposed here so as to discriminate between different radiative systems according to their complexity? Or, in other words, among  radiative systems is there a simplest one ?  Obviously this would require a closer examination of the orders higher than the leading ones in the complexity factors.
\item Is it possible to discriminate between different static spacetimes? Again, this would require to go beyond the orders employed here.
\end{itemize}
\begin{widetext}
\begin{table}[htp]
\caption{Complexity factors for different spacetimes of the Bondi metric}
\begin{center}
COMPLEXITY HIERARCHY
 
\begin{tabular}{ccccccccccccccccc}
\hline\hline
$complex. fac. \diagdown    spacetimes$ &$\quad$&
Minkowski &$\quad$ &
Static &$\quad$ &
NNNRS &$\quad$ &
NNRS &$\quad$ &
Radiative &$\quad$ &
 \vspace{0.3cm} \\  \hline
${\cal E}_1$&$\quad$ &0  &$\quad$ &  0     &$\quad$ & 0   &$\quad$ &0 &$\quad$ & ${\cal E}^{(n)}_1\neq 0$, $n\geq 1$\\ \hline
${\cal E}_2$ &$\quad$ & 0  & &     ${\cal E}^{(3)}_2=3M$& & ${\cal E}^{(3)}_2=3M$&  & ${\cal E}^{(3)}_2=3M+\frac{M_{\theta \theta}}{4}-\frac{M_{ \theta}\cot \theta}{4}$&$\quad$ & ${\cal E}^{(n)}_2\neq 0$, $n\geq 1$ \\ \hline
${\cal E}_3$&$\quad$ & 0  &$\quad$ &   0       &$\quad$ & 0  &$\quad$ & ${\cal E}^{(3)}_3=\frac{1}{2}(M_{\theta \theta}-M_{ \theta}\cot \theta)$&$\quad$ & ${\cal E}^{(n)}_3\neq 0$, $n\geq 1$ \\ \hline
\end{tabular}
\end{center}
\label{data}
\end{table}
\end{widetext}
where ${\cal E}^{(n)}_{1,2,3}$ are de coefficients of order $\mathcal{O}(r^{-n})$.

\begin{widetext}
\begin{table}[htp]
\caption{The magnetic parts of the Weyl tensor and the vorticity for different spacetimes of the Bondi metric}
\begin{center}
Magnetic parts and vorticity
 
\begin{tabular}{ccccccccccccccccc}
\hline\hline
$Mag. Weyl; \Omega. \diagdown    spacetimes$ &$\quad$&
Minkowski &$\quad$ &
Static &$\quad$ &
NNNRS &$\quad$ &
NNRS &$\quad$ &
Radiative &$\quad$ &
 \vspace{0.3cm} \\  \hline
$H_1$&$\quad$ &0  &$\quad$ &  0     &$\quad$ & 0   &$\quad$ &0 &$\quad$ & $H^{(n)}_1\neq 0$, $n\geq 1$\\ \hline
$H_2$ &$\quad$ & 0  &$\quad$ &   $0$      &$\quad$ & $0$   &$\quad$ & $H^{(3)}_2=-\frac{1}{4}(M_{\theta \theta}-M_{ \theta}\cot \theta)$ &$\quad$ & $ H^{(n)}_2\neq 0$, $n\geq 1$  \\ \hline
$\Omega$&$\quad$ & 0  &$\quad$ &   0       &$\quad$ & 0  &$\quad$ & $\Omega^{(2)}=M_{ \theta}$&$\quad$ & $\Omega^{(n)}\neq 0$, $n\geq 1$ \\ \hline
\end{tabular}
\end{center}
\label{data}
\end{table}
\end{widetext}

\appendix 
\section{The complexity factors}
\begin{widetext}
\begin{eqnarray}
{\cal E}_1=\frac{1}{r^2}\left(2 c_u \cot{\theta}+c_{\theta u}\right) +\mathcal{O}(r^{-n}), \quad n\geq 4,
\label{e1}
\end{eqnarray}

\begin{eqnarray}
{\cal E}_2=\frac{1}{r}c_{uu}
-\frac{1}{2r^2}\left(c_{\theta \theta u} -4 M c_{u u}+2c_{u}+c_{\theta u}\cot{\theta}-\frac{4 c_{u}}{\sin^2{\theta}}\right)\nonumber \\
+\frac{1}{r^3}\left[c c_u +2 c_\theta c_{\theta u}+3M 
+\frac{\cot{\theta}}{2}\left(3 c_u c_\theta + 5 c c_{\theta u}\right) -M_u c + \frac{1}{2}M_{\theta \theta}+N_{\theta u} 
 +P_{uu}\right. \nonumber \\ 
 \left.- \cot{\theta}\left(M c_{\theta u}+\frac{1}{2}M_\theta +N_u -N c_{uu}\right)
 -M c_u\left(1-\frac{4}{\sin^2{\theta}}\right)+c_u\left(c c_u +\frac{1}{2}c_{\theta \theta}\right)
\right. \nonumber \\ 
\left. +c_{uu}\left(4M^2+N_{\theta}\right)-c_{\theta \theta u}\left(M-\frac{3}{2}c\right)\right] +\mathcal{O}(r^{-n}), \quad n\geq 4,
\label{e2}
\end{eqnarray}

\begin{eqnarray}
{\cal E}_3=\frac{2}{r}c_{uu}
-\frac{1}{r^2}\left(c_{\theta \theta u} -4 M c_{u u}+2c_{u}+c_{\theta u}\cot{\theta}-\frac{4 c_{u}}{\sin^2{\theta}}\right)
\nonumber \\
+\frac{1}{r^3}\left[-4 c c_u +4 c_\theta c_{\theta u} 
 +\cot{\theta}\left(3c_u c_\theta + 5c c_{\theta u}\right) -2M_u c + M_{\theta \theta}+2N_{\theta u} +2P_{uu}\right.\nonumber \\
\left. -\cot{\theta}\left(2M c_{\theta u}+M_\theta +2N_u -2N c_{uu}\right)
 -2M c_u \left(1-\frac{4}{\sin^2{\theta}}\right)+c_u\left(2c c_u +c_{\theta \theta}\right)\right. \nonumber\\
\left. +2c_{uu}\left(4M^2+N_{\theta}\right)-c_{\theta \theta u}\left(2M-3c\right)\right] +\mathcal{O}(r^{-n}), \quad  n\geq 4,
\label{e3}
\end{eqnarray}
\end{widetext}

\section{The magnetic part of the Weyl tensor}
\begin{eqnarray}
H_1=-\frac{1}{r^2} \left(2 c_u \cot{\theta}+c_{\theta u}\right) +\mathcal{O}(r^{-n}), \quad n\geq 4,
\label{h1}
\end{eqnarray}
\\
\begin{widetext}
\begin{eqnarray}
H_2=-\frac{1}{r}c_{uu}-\frac{1}{r^2}\left[-c_u\left(1-\frac{2}{\sin^2{\theta}}\right)
 -\frac{\cot{\theta}}{2}c_{\theta u}+ 2 c_{uu}(M-c)-\frac{1}{2}c_{\theta \theta u}\right]\nonumber\\
-\frac{1}{r^3}\left\{-Mc_u\left(1-\frac{4}{\sin^2{\theta}}\right) - \frac{4cc_u}{\sin^2{\theta}}
+\cot{\theta}\left[\frac{3}{2}c_u c_\theta - N_u-\frac{1}{2}M_{\theta}+N c_{uu}+\left(\frac{7}{2}c-M\right)c_{\theta u}\right]+\left(\frac{5}{2}c-M\right)c_{\theta \theta u}\right. \nonumber \\
\left.+\frac{1}{2}c_{\theta \theta} c_{u}+2c_{\theta}c_{\theta u}+c c_{u}^2+\frac{1}{2}M_{\theta \theta}-cM_{u}
+N_{\theta u}+P_{uu}+c_{uu}\left(4c^2+4M^2-4Mc+N_{\theta}\right)\right\}+\mathcal{O}(r^{-n}),\quad  n\geq 4\nonumber\\
\label{h2}
\end{eqnarray}
\end{widetext}

where $P=C-\frac{c^3}{6}$.
\section{The vorticity}
\begin{widetext}
\begin{eqnarray}
\omega^{\phi}  =  -\frac{e^{-2\beta}}{2 r^2 \sin{\theta}} \left[ 2
\beta_\theta e^{2\beta} - \frac{2 e^{2\beta} A_\theta}{A}
- \left(U r^2 e^{2\gamma}\right)_r  
  +  \frac{2 U r^2 e^{2\gamma}}{A} A_r +
\frac{e^{2\beta}\left(U r^2 e^{2\gamma}\right)_u}{A^2} - \frac{Ur^2
e^{2\gamma}}{A^2} 2 \beta_u e^{2\beta} \frac{}{} \right], \label{om3}
\end{eqnarray}
\end{widetext}
and for the absolute value of $\omega^\alpha$ we get
\begin{widetext}
\begin{eqnarray}
\Omega  \equiv  \left(- \omega_\alpha \omega^\alpha\right)^{1/2} =
  \frac{e^{-2\beta -\gamma}}{2 r}
\left[2 \beta_\theta e^{2\beta} - 2 e^{2\beta} \frac{A_\theta}{A}
  -  \left(U r^2 e^{2\gamma}\right)_r
+  2 U r^2 e^{2\gamma} \frac{A_r}{A}
  +  \frac{e^{2\beta}}{A^2} \left(U r^2 e^{2\gamma}\right)_u
  - 2 \beta_u \frac{e^{2\beta}}{A^2} U r^2 e^{2\gamma}
\right] \label{OM}
\end{eqnarray}
\end{widetext}
Feeding back (\ref{ga}--\ref{be}) into (\ref{OM}) and
keeping only the two leading terms, we obtain
\begin{widetext}
\begin{eqnarray}
\Omega  = -\frac{1}{2r} ( c_{u \theta}+2 c_u \cot \theta)  +\frac 1{r^2} \left[ M_{\theta}-M (c_{u \theta}+2 c_u \cot 
\theta)-c c_{u
\theta}+6 c c_u \cot \theta+2 c_u c_{\theta} \right].
\label{Om2}
\end{eqnarray}
\end{widetext}

\begin{acknowledgments}
This  work  was partially supported by the Spanish  Ministerio de Ciencia e
Innovaci\'on under Research Projects No.  FIS2015-65140-P (MINECO/FEDER).
L.H. thanks      Departament de F\'isica at the  Universitat de les  Illes Balears, for financial support and hospitality. ADP  acknowledges hospitality of the
Departament de F\'isica at the  Universitat de les  Illes Balears. J. C.  gratefully acknowledges financial support from the Spanish Ministerio de Econom\'ia y Competitividad through grant ref.: FPA2016-76821.
\end{acknowledgments}

\end{document}